\documentclass{article}
\usepackage{waspaa23,amsmath,graphicx,url,times}
\usepackage{color}

\usepackage{tikz}
\usepackage{tkz-graph}
\usetikzlibrary{positioning}
\usetikzlibrary{bayesnet}
\usetikzlibrary {matrix}
\usepackage{twoopt}
\usepackage{xparse,xfp}
\usepackage{color, colortbl}
\usepackage{amssymb}
\usepackage{multirow}
\usepackage{multicol}
\usepackage{booktabs}
\usepackage{varwidth}
\usepackage{url}

\usepackage{ifthen}
\title{Unsupervised Improvement of Audio-Text Cross-Modal Representations}

\name{\begin{varwidth}{16cm}\begin{center} {Zhepei Wang}$^\sharp$, {Cem Subakan}$^{\flat \natural \chi}$,  
     {Krishna Subramani}$^\sharp$, {Junkai Wu}$^\sharp$, Tiago Tavares$^\dagger$, \\ Fabio Ayres$^\dagger$, Paris Smaragdis$^{\sharp\alpha}$
     \end{center}
     \end{varwidth}
    } 
\address{
     $^\sharp$University of Illinois at Urbana-Champaign, $^\flat$Université Laval, $^\natural$Concordia University, \\ $^\chi$Mila-Quebec AI Institute, $^\dagger$Insper, $^\alpha$Amazon Web Services}

\begin{document}

\ninept
\maketitle

\begin{sloppy}

\begin{abstract}
  

  

Recent advances in using language models to obtain cross-modal audio-text representations have overcome the limitations of conventional training approaches that use predefined labels. This has allowed the community to make progress in tasks like zero-shot classification, which would otherwise not be possible. However, learning such representations requires a large amount of human-annotated audio-text pairs. In this paper, we study unsupervised approaches to improve the learning framework of such representations with unpaired text and audio. We explore domain-unspecific and domain-specific curation methods to create audio-text pairs that we use to further improve the model. We also show that when domain-specific curation is used in conjunction with a soft-labeled contrastive loss, we are able to obtain significant improvement in terms of zero-shot classification performance on downstream sound event classification or acoustic scene classification tasks. 
  
\end{abstract}

\begin{keywords}
Audio-text representation learning, data augmentation, contrastive learning, sound event classification, acoustic scene classification
\end{keywords}
\newcommand{\R}{\mathbb{R}}
\newcommand{\mone}{\mathbbm{1}}
\newcommand{\bzero}{\mathbf{0}}
\newcommand{\bD}{\mathbf{D}}
\newcommand{\bH}{\mathbf{H}}
\newcommand{\bM}{\mathbf{M}}
\newcommand{\bS}{\mathbf{S}}
\newcommand{\bU}{\mathbf{U}}
\newcommand{\bV}{\mathbf{V}}
\newcommand{\bW}{\mathbf{W}}
\newcommand{\bZ}{\mathbf{Z}}
\newcommand{\ba}{\mathbf{a}}
\newcommand{\bh}{\mathbf{h}}
\newcommand{\bq}{\mathbf{q}}
\newcommand{\bs}{\mathbf{s}}
\newcommand{\bv}{\mathbf{v}}
\newcommand{\bx}{\mathbf{x}}
\newcommand{\by}{\mathbf{y}}
\newcommand{\bz}{\mathbf{z}}
\newcommand{\calD}{\mathcal{D}}
\newcommand{\calE}{\mathcal{E}}
\newcommand{\calL}{\mathcal{L}}
\newcommand{\cat}{\text{Concat}}

\newcommand{\zp}[1]{#1}
\newcommand{\cem}[1]{#1}

\section{Introduction}
\label{sec:intro}

Representation learning methods such as Self-Supervised Learning (SSL) \cite{gui2023survey} expand the limited scope of supervised learning by learning representations that can be applied to a large variety of downstream tasks. However, the mainstream SSL methods in the literature typically train an encoder on uni-modal data \cite{simclr, bert, ssast}.  

Learning cross-modal representations that involve text adds the additional flexibility of incorporating language in downstream tasks. This enables downstream tasks such as zero-shot classification possible, where the model is able to perform classification without being restricted by a pre-defined and explicitly annotated label set. The cross-modal representations can also be used in other tasks, such as audio-to-text and text-to-audio retrieval.

Learning cross-modal representations has been explored in computer vision under prior works including CLIP \cite{clip}, Florence \cite{yuan2021florence}, and ALIGN \cite{jia2021scaling}. AudioCLIP \cite{guzhov2021audioclip} extends the CLIP framework to the audio domain by incorporating an audio encoder to learn a joint embedding space for audio, vision, and language using aligned data across the three domains. CLAP~\cite{clap}, on the other hand, learns audio-text embeddings directly without depending on the image domain. It aims to maximize the similarity of the text-and-audio representations that correspond to the paired audio and caption within a given batch.
To train high-quality representations, these methods require a large number of paired audio and text items. {While in-the-wild audio-text pairs exist at an extensive scale (e.g., captions from online video, metadata from audio datasets), the text is likely to be irrelevant to the sound events presented in the audio and is therefore unsuitable for training audio representations.} Collecting \cem{high quality} captioned audio is \cem{therefore} an expensive task, hence data availability might constitute a bottleneck for scaling the audio-text pretraining. To overcome the limitation, Wav2CLIP \cite{wu2022wav2clip} uses audio-image pairs from video clips to learn audio-text correspondence by distilling from the pre-trained CLIP model. VIP-ANT \cite{vipant} extends Wav2CLIP by mining additional audio-text pairs from video and text data using pre-trained CLIP. The LAION-CLAP \cite{laionclap2023} augments the training data by performing keyword-to-caption generation from tags or labels of audio clips using a pre-trained language model. These approaches, however, assume the existence of either an additional anchor modality with paired annotations or a pre-trained model for generating text.




In this work, we explore the possibility of improving the zero-shot classification performance of audio-text representations using unpaired text and audio. For this purpose, we first train an initial teacher model using paired audio clips and captions; we then use this teacher model to automatically align textual descriptions to in-the-wild audio files, using the pairs aligned with higher confidence to train a new model.
We then argue that this performance can be further improved by curating a refined, domain-specific dataset that is more akin to the zero-shot classification domain.


\footnote{Implementation available at \url{https://github.com/zhepeiw/clap_curation}}
Our contributions in this paper are as follows,
\begin{itemize}
    \setlength\itemsep{.003cm}
    \item We propose domain-unspecific and domain-specific curation methods and show the improvement on three downstream zero-shot audio classification tasks.

    \item We propose a soft-labeled training objective that can avoid learning with hard labels in cases where the batch contain similar data items. We show that this training objective significantly improves the performance under domain-specific curation strategies. 

    \item We show that the proposed curation methods significantly improve the model even in the case where the teacher model is trained on a small amount of paired data (\%10 of the data). 
\end{itemize}

\section{Methodology}
\label{sec:methodology}

\tikzstyle{block} = [draw, fill=lightgray, rectangle, 
    minimum height=3em, minimum width=4em]

\subsection{Contrastive Language-Audio Pretraining and Zero-Shot Classification}
\label{ssec:method_clap}
CLAP, ``Contrastive Language-Audio Pretraining'', learns a joint latent space between text and audio by maximizing the similarity between the text and audio latent representation for the text caption and its corresponding audio signal. Let, $X_t, X_a$ respectively denote a batch of text and audio. In the CLAP model, the latent representation is obtained by passing the text and audio through the text and audio encoders $f_t(.)$, and $f_a(.)$ such that ${L_t} = f_t(X_t), \; L_a =  f_a(X_a)$, 
where $L_t \in \mathbb R^{N \times T}$, $L_a \in \mathbb R^{N \times A}$, such that $T$ is the latent dimensionality of text, $A$ is the latent dimensionality of audio, and $N$ is the batch size. CLAP trains a joint latent space by passing $L_t$ and $L_a$ through fully-connected layers such that ${t} = \textbf{MLP}_t(L_t), \; {a} = \textbf{MLP}_a(L_a)$,
where $\textbf{MLP}(.)$ denotes the multi-layer perceptron transformation layers, $t \in \mathbb R^{N \times d}$, and $a \in \mathbb R^{N \times d}$ respectively denote the latent variables with same latent dimensionality $d$. The model then tries to maximize the diagonal entries on the matrix $C= t a^\top$. This translates into the following training loss function,  
\begin{align}
    \mathcal L(C) =   \frac{1}{2}\sum_{i=1}^N \Bigl( \log({\textbf{Softmax}_t(C/\tau)_{i,i}}) + \log({\textbf{Softmax}_a(C/\tau)_{i,i}}) \Bigr),
    \label{eq:clap}
\end{align}
where $\textbf{Softmax}_t(.)$ and $\textbf{Softmax}_a(.)$ respectively denote Softmax functions along text and audio dimensions, $\tau$ is a learnable temperature scaling parameter, and the $C_{i,i}$ denotes the diagonal elements of the $C$ matrix. We show the training forward pass pipeline in the left panel of Figure \ref{fig:clap}. 

\begin{figure*}[ht]
    \centering
            \resizebox{6.5cm}{!}{
    		\begin{tikzpicture}[ampersand replacement=\&]
			\node [] (1) {\color{red}{\texttt{INPUT TEXT}}};

			\node [block, right of=1, xshift=2cm] (model) {Text Encoder}; 
			\node [right of=model, xshift=1.9cm, yshift=.0cm] (2) {}; 
			\node [block, below of=model, xshift=0cm, yshift=-1cm] (embed) {Audio Encoder}; 
			\node [left of=embed, xshift=-2cm] (voice) {\includegraphics[scale=0.2, trim=3cm 2.4cm 4cm 2cm, clip]{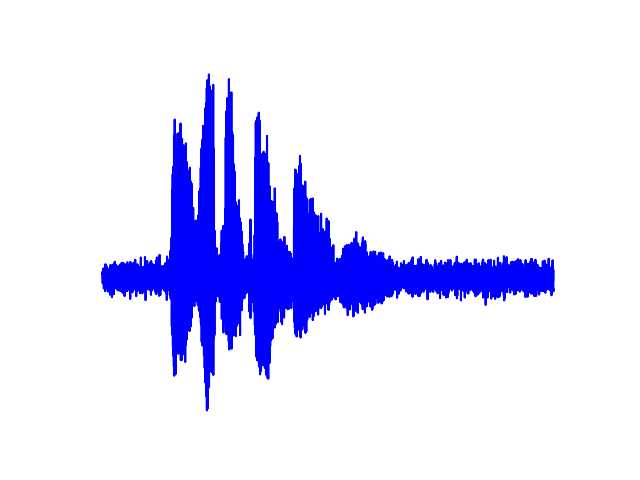}}; 

			\matrix (mat1) [right of=2, xshift=-.4cm, nodes={fill=blue!20,minimum size=5mm}]
			  {
				  \node [fill=cyan] {$t_1^\top a_1$}; \& \node{$t_1^\top a_2$}; \& \node {$t_1^\top a_3$}; \\
			    \node   {$t_2^\top a_1$}  ; \& \node [fill=cyan] {$t_2^\top a_2$}; \& \node {$t_2^\top a_3$} ; \\
			    \node {$t_3^\top a_1$}  ; \& \node {$t_3^\top a_2$} ; \& \node [fill=cyan] {$t_3^\top a_3$} ; \\
			  };

			\matrix (mat2) [below of=mat1, yshift=-.25cm, nodes={fill=green!20,minimum size=5mm}]
			  {
				  \node {$a_1$}; \& \node (mb) {$a_2$}; \& \node {$a_3$}; \\
		          };
			\matrix (mat3) [left of=mat1, xshift=-.7cm, nodes={fill=yellow!20,minimum size=5mm}]
			  {
		            \node {$t_1$}; \\ 
			    \node (t2) {$t_2$}; \\ 
			    \node {$t_3$}; \\
		          };

			\draw [->] (1) -- (model);
			\draw [->] (model) -- (t2);
			\draw [->] (voice) -- (embed);
			\draw [->] (embed) -| (mb);

		\end{tikzpicture}
        }
        \resizebox{6.5cm}{!}{
        \begin{tikzpicture}[ampersand replacement=\&]
			\node [] (1) {\color{red}{\texttt{This is a DOG sound}}}; 
            \node [above of=1, yshift=-.5cm] (1above) {\color{red}{\texttt{This is a CAT sound}}}; 
            \node [below of=1, yshift=+.5cm] (1above) {\color{red}{\texttt{This is a BIRD sound}}};

			\node [block, right of=1, xshift=2.5cm] (model) {Text Encoder}; 
			\node [right of=model, xshift=1.9cm, yshift=.0cm] (2) {}; 
			\node [block, below of=model, xshift=0cm, yshift=-1cm] (embed) {Audio Encoder}; 
			\node [left of=embed, xshift=-2cm] (voice) {\includegraphics[scale=0.2, trim=3cm 2.4cm 4cm 2cm, clip]{timeseries_wav.png}}; 

			\matrix (mat1) [right of=2, xshift=-1.4cm, nodes={fill=blue!20,minimum size=5mm}]
			  {
				  \node [fill=cyan] {$t_1^\top a_\text{test}$};  \\
			      \node {$t_2^\top a_\text{test}$} ; \\
			    \node {$t_3^\top a_\text{test}$} ; \\
			  };

			\matrix (mat2) [below of=mat1, yshift=-.25cm, nodes={fill=green!20,minimum size=5mm}]
			  {
				    \node (mb) {$a_\text{test}$};  \\
		          };
			\matrix (mat3) [left of=mat1, xshift=.2cm, nodes={fill=yellow!20,minimum size=5mm}]
			  {
		            \node {$t_1$}; \\ 
			    \node (t2) {$t_2$}; \\ 
			    \node {$t_3$}; \\
		          };

			\draw [->] (1) -- (model);
			\draw [->] (model) -- (t2);
			\draw [->] (voice) -- (embed);
			\draw [->] (embed) -| (mb);

		\end{tikzpicture}
        }
  
    \caption{\textbf{(left)} The training of the CLAP model for learning cross-modal representations. \textbf{(right)} Zero shot classification with the CLAP model}
    \label{fig:clap}
    \vspace{-.5cm}
\end{figure*}

The CLAP model is able to perform zero-shot classification by simply calculating the similarity of a given audio to a fixed set of text prompts constructed from class labels. That is, the classification decision is simply taken to be $\widehat c = \arg \max_j t^\top_j a_\text{test},$
where $\widehat c$ is the zero-shot classification decision, $a_\text{test}$ is the embedding for the test audio, and $t_j$ is the text embedding corresponding to the label of class $j$. We show the pipeline of zero-shot classification in the right panel of Figure \ref{fig:clap}. 

\subsection{Unsupervised Improvement of Cross-Modal Audio-Text Representations}
\label{ssec:method_ucur}
In this section, we describe the different strategies we employ to curate a paired dataset from unpaired text and audio. This curated dataset is used to train a student model in order to improve upon the zero-shot performance of the teacher model. We call this curated dataset the \emph{Improvement-Set}. We follow two different strategies to curate this dataset. We call the first strategy the ``Domain-Unspecific'' (DU) dataset curation, where we form pairs using in-the-wild audio and text. The second strategy is ``Domain-Specific'' (DS) where we explicitly try to find audio recordings that are more relevant to the zero-shot classification task at hand. 

\subsubsection{Domain-Unspecific Improvement-Set Curation}
\label{sssec:method_du}
For this strategy, we use a teacher CLAP model to curate an Improvement-Set. We match the captions of the training data with audio from a large dataset such as AudioSet \cite{audioset}. We compute the cosine similarity between each pair of audio and text embeddings, and keep pairs with similarity above a threshold $\sigma \in [0, 1]$. We show this strategy in Figure \ref{fig:DU}. As we showcase in the experiments, this strategy provides an improvement over the base model; however, with a domain-specific refinement of the Improvement-Set, we can further enhance the zero-shot performance. 

\tikzstyle{ellipseblk} = [draw, fill=lightgray, ellipse, 
    minimum height=3em, minimum width=4em]

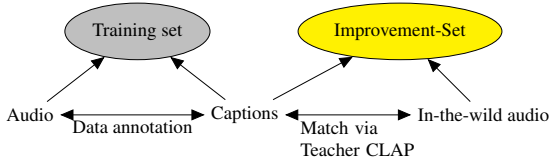
\begin{figure}
    \resizebox{7.51cm}{!}{
    \centering
    \begin{tikzpicture}
			\node [ellipseblk] (trset) {Training set};
            \node [below of=1, xshift=-1.8cm, yshift=-.4cm] (audio) {Audio};
            \node [below of=1, xshift=+1.8cm, yshift=-.4cm] (capt) {Captions};

            \node [ellipseblk, fill=yellow, right of=trset, xshift=3.4cm] (IS) {Improvement-Set};
            \node [below of=IS, xshift=1.4cm, yshift=-.4cm] (iwa) {In-the-wild audio};

            \draw [->] (audio) -- (trset);
            \draw [->] (capt) -- (trset);
            \draw [<->] (capt) -- node [yshift=-.4cm, xshift=.2cm, text width=2cm] (match) {Match via Teacher CLAP} (iwa);
            \draw [<->] (audio) -- node [yshift=-.2cm] (match) {Data annotation} (capt);

            \draw [->] (capt) -- (IS);
            \draw [->] (iwa) -- (IS); 
            
	\end{tikzpicture}
    }
    \caption{Domain-Unspecific (DU) Curation of Improvement-Set. {With the pre-trained teacher CLAP model, the embedding of each in-the-wild audio clip is matched against the embeddings of all training captions. The new audio-text pairs with similarity above a certain threshold $\sigma$ are included in the Improvement-Set.}}
    \label{fig:DU}
    \vspace{-.6cm}
\end{figure}

\subsubsection{Domain-Specific Improvement-Set Curation}
\label{sssec:method_ds}
For this strategy, the main idea is to narrow down the captions used for the Improvement-Set by calculating text-to-text similarities between the captions from the training set and in-domain labels for a zero-shot classification task. Once these similarities are obtained, the first domain-specific (DS) strategy is to narrow down the training set by picking the subset of audio-text pairs that are most related to the downstream task through the text modality. We illustrate this procedure in Figure \ref{fig:DS}. 

Another option is to augment the domain-specific Improvement-Set by involving in-the-wild audio data. The process to create the Augmented Domain-Specific (ADS) Improvement-Set, as demonstrated in Figure \ref{fig:ADS}, is as follows: 

\begin{enumerate}
     \setlength\itemsep{.003cm}
    \item We calculate the similarities between the text embeddings of class labels from the in-domain dataset and the captions from the teacher's training set. We take the most similar captions from this set by thresholding the similarity values. 
    \item We find the correspondences between the domain-specific captions and the in-the-wild audio using the teacher CLAP. 
\end{enumerate}

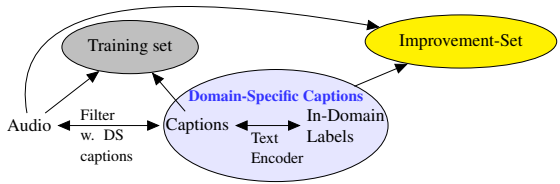
\begin{figure}[h!]
    \centering
    \resizebox{7.51cm}{!}{
    \begin{tikzpicture}
    
            \draw [fill=blue!10] (1.6,-1.4) ellipse (2cm and 1cm);
			\node [ellipseblk, xshift=-1cm] (trset) {Training set};
            \node [below of=trset, xshift=-1.8cm, yshift=-.4cm] (audio) {Audio};
            \node [below of=trset, xshift=+1.2cm, yshift=-.4cm] (capt) {Captions
            };
            \node [right of=capt, xshift=+1.7cm, text width=1.5cm] (IDC) {In-Domain Labels};


            \draw [->] (audio) -- (trset);
            \draw [->] (capt) -- (trset);
            \draw [<->] (audio) -- node [yshift=-.2cm, text width=1cm] (match) {\footnotesize{Filter w. DS captions}} (capt);
            \draw [<->] (capt) -- node [yshift=-.4cm, xshift=.2cm, text width=1cm] (match) {\footnotesize{Text Encoder}} (IDC);
            \node [above of=trset, xshift=2.6cm, yshift=-1.9cm] (DSCapt) {\color{blue!80}{\footnotesize{\textbf{Domain-Specific Captions}}}};
            
            \node [ellipseblk, fill=yellow, above of=DSCapt, xshift=3.3cm, yshift=-.0cm] (IS) {Improvement-Set};
            \draw [->] (3.0, -0.7) -- (IS);
            \draw [->] (audio) to[out=100, in=170]  (IS);
	\end{tikzpicture}
    }
    \caption{Domain-Specific (DS) Curation of Improvement-Set with Domain-Specific Audio. {The Improvement-Set consists of audio-caption pairs from the training set that are most relevant to the downstream task by measuring the similarity between the embeddings of the training captions and in-domain labels of the downstream task.}}
    \label{fig:DS}
\end{figure}

\begin{figure}[h!]
    \centering
    \resizebox{8.51cm}{!}{
    \begin{tikzpicture}
    
            \draw [fill=blue!10] (1.6,-1.4) ellipse (2cm and 1cm);
			\node [ellipseblk, xshift=-1cm] (trset) {Training set};
            \node [below of=trset, xshift=-1.8cm, yshift=-.4cm] (audio) {Audio};
            \node [below of=trset, xshift=+1.2cm, yshift=-.4cm] (capt) {Captions
            };
            \node [right of=capt, xshift=+1.7cm, text width=1.5cm] (IDC) {In-Domain Labels};

            \node [ellipseblk, fill=yellow, right of=trset, xshift=5.9cm] (IS) {Improvement-Set};
            \node [below of=IS, xshift=1.0cm, yshift=-.4cm] (iwa) {In-the-wild audio};


            \draw [->] (audio) -- (trset);
            \draw [->] (capt) -- (trset);
            \draw [<->] (IDC) -- node [yshift=-.4cm, xshift=.2cm, text width=2cm] (match) {Match via Teacher CLAP} (iwa);
            \draw [<->] (audio) -- node [yshift=-.2cm, text width=2.0cm] (match) {\footnotesize{Data annotation}} (capt);
            \draw [<->] (capt) -- node [yshift=-.4cm, text width=1.0cm, xshift=.2cm] (match) {\footnotesize{Text Encoder}} (IDC);
            \node [above of=trset, xshift=2.6cm, yshift=-1.9cm] () {\color{blue!80}{\footnotesize{\textbf{Domain-Specific Captions}}}};
            
            \draw [->] (iwa) -- (IS); 
            \draw [->] (3.2, -.8) -- (IS);
	\end{tikzpicture}
    }
    \caption{Augmented Domain-Specific (ADS) Curation of Improvement-Set. {First, DS Curation is performed to obtain the subset of captions from the training set that are most related to the downstream task. Then, in-the-wild audio clips are aligned with this subset of captions with the pre-trained teacher CLAP model.}}
    \label{fig:ADS}
\end{figure}
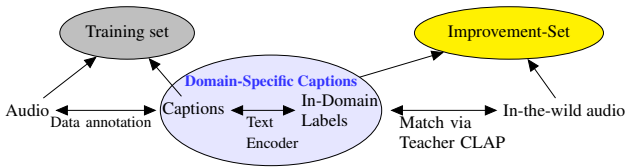

\subsection{Using Soft Labels in the CLAP Loss}
\label{ssec:method_sl}

An underlying issue with the original CLAP learning objective in Equation \eqref{eq:clap} is that it neglects local similarities for data within the same batch and treats all negative samples equally. It is possible to sample audio-text pairs with similar content from the same batch (i.e, multiple clips of ``dog-barking''), while the objective in Equation \eqref{eq:clap} penalizes the model for not discriminating these similar instances. This issue is exacerbated when the training data is less diverse. For instance, with the DS or ADS Improvement-Set, it is more likely to sample similar inputs within a batch. 

The original learning framework of CLAP is equivalent to solving an $N$-class classification problem for a batch size of $N$, where the label for the $i$-th sample of each batch is the one-hot vector $\by_i \in \R^N$ with only the $i$-th entry equal to 1. Similar to prior studies in image-text pretraining \cite{clip_self_distill, softclip}, we propose a label-softening technique when training the student model. The soft labels are obtained from the similarity between input samples within each batch, as illustrated in Figure \ref{fig:SL}.
For each input batch,
we obtain the soft targets by estimating the intra-modal similarity using the teacher audio ($\tilde{a}$) and text ($\tilde{t}$) embeddings.
We define,
\begin{align}
    \Tilde{C}_a = \Tilde{a} \Tilde{a}^\top , \; \Tilde{C}_t = \Tilde{t} \Tilde{t}^\top,
\end{align}
where $\Tilde{C}_a, \Tilde{C}_t \in \R^{N\times N}$ represent the intra-batch, intra-modal similarity matrices for audio and text, respectively. For the $i$-th sample in this batch, we define the audio-to-text soft label as,
\begin{align}
     \Tilde{\by}_{a \rightarrow t}^{(i)} = (1 - \beta)\by_i + \beta \Tilde{C}_a^{(i)},
     \label{eq:a2t_softlabel}
\end{align}
where $\Tilde{C}_a^{(i)}$ is the $i$-th row of the matrix $\Tilde{C}_a$ and the coefficient \\ $\beta \in [0, 1]$ adjusts the weights of the hard and soft labels; in symmetry, the text-to-audio soft label can be formulated as, 
\begin{align}
     \Tilde{\by}_{t \rightarrow a}^{(i)} = (1 - \beta)\by_i + \beta \Tilde{C}_t^{(i)}.
     \label{eq:t2a_softlabel}
\end{align}
The overall learning objective is therefore, \\ $\Tilde{\mathcal{L}}(C) = \frac{1}{2} (\Tilde{\calL}_t(C) + \Tilde{\calL}_a(C))$, with
\begin{align}
    \begin{split}
        \Tilde{\calL}_t(C) & = \frac{1}{N} \sum_{i=1}^N \calD_{KL}( \Tilde{\by}_{a \rightarrow t}^{(i)} \| \textbf{Softmax}_t(C/\tau)^{(i)} ), \\
         \Tilde{\calL}_a(C) & = \frac{1}{N} \sum_{i=1}^N \calD_{KL}( \Tilde{\by}_{t \rightarrow a}^{(i)} \| \textbf{Softmax}_a(C/\tau)^{(i)}), 
    \end{split}
    \label{eq:sl_kldiv}
\end{align}
where $\calD_{KL}$ denotes KL-Divergence, and $C$ denotes the similarity matrix obtained from the student model. 

\tikzstyle{sumt}   = [circle, minimum width=8pt, draw, inner sep=0pt, path picture={\draw (path picture bounding box.east) -- (path picture bounding box.west) (path picture bounding box.south) -- (path picture bounding box.north);}]

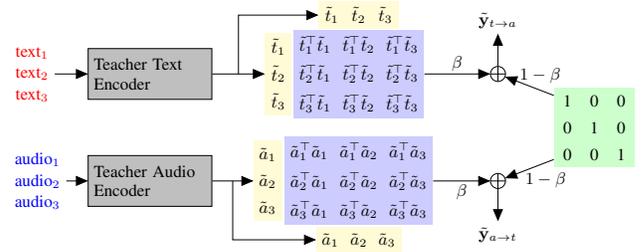
\begin{figure}
    \centering
    \resizebox{8.5cm}{!}{
    		\begin{tikzpicture}[ampersand replacement=\&]
			\node [] (1) {\color{red}{$\text{text}_2$}}; 
            \node [above of=1, yshift=-.6cm] (text1) {\color{red}{$\text{text}_1$}}; 
            \node [below of=1, yshift=+.6cm] (text3) {\color{red}{$\text{text}_3$}}; 

			\node [block, right of=1, xshift=1.2cm, text width=2.1cm] (model) {Teacher Text \\Encoder}; 
			\node [right of=model, xshift=1.9cm, yshift=.0cm] (2) {}; 
			\node [block, below of=model, xshift=0cm, yshift=-1cm, text width=2.1cm] (embed) {Teacher Audio \\Encoder}; 
   
			\node [left of=embed, xshift=-1.1cm] (voice) {$\color{blue}{\text{audio}}_2$}; 
           \node [above of=voice, xshift=0cm, yshift=-.6cm] (voice1) {$\color{blue}{\text{audio}}_1$}; 
           \node [below of=voice, xshift=0cm, yshift=+.6cm] (voice3) {$\color{blue}{\text{audio}}_3$}; 

			\matrix (mat1) [right of=2, xshift=+0.0cm, yshift=.0cm, nodes={fill=blue!20,minimum size=5mm}]
			  {
				  \node  {$\tilde t_1^\top \tilde t_1$}; \& \node{$\tilde t_1^\top \tilde t_2$}; \& \node {$\tilde t_1^\top \tilde t_3$}; \\
			    \node   {$\tilde t_2^\top \tilde t_1$}  ; \& \node {$\tilde t_2^\top \tilde t_2$}; \& \node (t23) {$\tilde t_2^\top \tilde t_3$} ; \\
			    \node {$\tilde t_3^\top \tilde t_1$}  ; \& \node {$\tilde t_3^\top \tilde t_2$} ; \& \node {$\tilde t_3^\top \tilde t_3$} ; \\
			  };

			\matrix (mat2) [above of=mat1, yshift=+.1cm, nodes={fill=yellow!20,minimum size=5mm}]
			  {
				  \node (t1) {$\tilde t_1$}; \& \node (mb) {$\tilde t_2$}; \& \node {$\tilde t_3$}; \\
		          };
			\matrix (mat3) [left of=mat1, xshift=-.5cm, nodes={fill=yellow!20,minimum size=5mm}]
			  {
		        \node {$\tilde t_1$}; \\ 
			    \node (t2) {$\tilde t_2$}; \\ 
			    \node {$\tilde t_3$}; \\
		          };

            \matrix (mat4) [right of=embed, xshift=+2.9cm, yshift=.0cm, nodes={fill=blue!20,minimum size=5mm}]
			  {
				  \node  {$\tilde a_1^\top \tilde a_1$}; \& \node{$\tilde a_1^\top \tilde a_2$}; \& \node {$\tilde a_1^\top \tilde a_3$}; \\
			    \node   {$\tilde a_2^\top \tilde a_1$}  ; \& \node {$\tilde a_2^\top \tilde a_2$}; \& \node (a32) {$\tilde a_2^\top \tilde a_3$} ; \\
			    \node {$\tilde a_3^\top \tilde a_1$}  ; \& \node {$\tilde a_3^\top \tilde a_2$} ; \& \node {$\tilde a_3^\top \tilde a_3$} ; \\
			  };
            \matrix (mat5) [below of=mat4, yshift=-.1cm, nodes={fill=yellow!20,minimum size=5mm}]
			  {
				  \node (a1) {$\tilde a_1$}; \& \node (mba) {$\tilde a_2$}; \& \node {$\tilde a_3$}; \\
		          };
            \matrix (mat6) [left of=mat4, xshift=-.7cm, nodes={fill=yellow!20,minimum size=5mm}]
			  {
		        \node {$\tilde a_1$}; \\ 
			    \node (a2) {$\tilde a_2$}; \\ 
			    \node {$\tilde a_3$}; \\
		          };

            \matrix (eye) [right of=mat4, xshift=+3.4cm, yshift=+1cm, nodes={fill=green!20,minimum size=5mm}]
			  {
				  \node (eye11) {1}; \& \node{0}; \& \node {0};  \\
			    \node   {0}  ; \& \node {1}; \& \node {0} ;  \\
			    \node (eye31) {0}  ; \& \node {0} ; \& \node {1};  \\
			  };

            \node [sumt, left of=eye, xshift=-.8cm, yshift=-1cm] (sum1) {}; 
            \node [sumt, left of=eye, xshift=-.8cm, yshift=+1cm] (sum2) {}; 

			\draw [->] (1) -- (model);
			\draw [->] (model) -- node (mt2) {} (t2);
            \draw [->] (mt2.center) |- (t1);
			\draw [->] (voice) -- (embed);
			\draw [->] (embed) -- node (ma2) {} (a2);
            \draw [->] (ma2.center) |- (a1);

            \draw [->]  (eye31) -- node [yshift=-.2cm, xshift=.3cm] (msum1) {$1-\beta$} (sum1);
            \draw [->] (a32) -- node [yshift=-.2cm] {$\beta$} (sum1);

            \draw [->] (eye11) -- node [yshift=.2cm, xshift=.2cm]{$1-\beta$} (sum2);
            \draw [->] (t23) -- node [yshift=.2cm] {$\beta$} (sum2); 

            \node [below of=sum1] (res1) {$ \Tilde{\by}_{a \rightarrow t}$} ;
            \node [above of=sum2] (res2) {$\Tilde{\by}_{t \rightarrow a}$};
            \draw [->] (sum1) -- (res1);
            \draw [->] (sum2) -- (res2);

		\end{tikzpicture}
        }

    \caption{Computation pipeline for soft-labeled loss function}
    \label{fig:SL}
\end{figure}

\section{Experimental Configurations}
\label{sec:exp}

\subsection{Training Data}
\label{ssec:exp_data}
For training the teacher CLAP, we follow the original paper \cite{clap} by using 128k paired audio and text captions from FSD50k \cite{FSD50K}, ClothoV2 \cite{clotho}, AudioCaps \cite{audiocaps}, and MACS \cite{macs}. For the unsupervised curation in Section \ref{ssec:method_ucur}, we use the captions from the teacher training set and recordings from the unbalanced training set from AudioSet \cite{audioset} by excluding recordings that are corrupted or contained in AudioCaps for the teacher training, which results in more than 1.9 million clips; notice that no label information from AudioSet is used during curation. We use a similarity threshold $\sigma=0.7$ for the domain-unspecific curation and a threshold between 0.6 and 0.75 for the domain-specific experiments. The numbers of audio-caption pairs of the curated datasets are outlined in Table \ref{tab:curation_size}.

During training, each input audio recording is resampled to $44.1$~kHz. If longer than $5$ seconds, a random $5$-second segment of the recording is chosen; if shorter, the recording is zero-padded. 


\begin{table}[]
\caption{Number of audio-caption pairs used in the curated Improvement-Set. DU represents the domain-unspecific curation; DS refers to the domain-specific curation containing the subset of the teacher training set relevant to each downstream task; ADS is the domain-specific curation from the in-the-wild audio.}
\vspace{.1cm}
\label{tab:curation_size}
\centering
\resizebox{8.5cm}{!}{
\begin{tabular}{c|c c|c c|c c}
 \multirow{2}{*}{DU} & \multicolumn{2}{c|}{\textbf{ESC-50}} & \multicolumn{2}{c|}{\textbf{UrbanSound8K}} & \multicolumn{2}{c}{\textbf{TUT17}}\\
& DS & ADS & DS & ADS & DS & ADS \\
\midrule
$5.0\!\times\!\!10^6$ & $15.6\!\times\!\!10^3$ & $1.30\!\times\!\!10^6$ & $9.7\!\times\!\!10^3$ & $1.1\!\times\!\!10^6$ & $8.7\!\times\!\!10^3$ & $0.3\!\times\!\!10^6$ \\
\end{tabular}
}
\end{table}

\subsection{Training Details}
\label{ssec:exp_train}


Following CLAP \cite{clap}, we use the CNN14 \cite{pann} as the audio encoder and the BERT \cite{bert} as the text encoder, each followed by a two-layer MLP as the projection layer. All modules are initialized following the setups in \cite{clap}.
The embedding vectors have a dimension of 1024. The temperature parameter $\tau$ is initialized to 0.007, and the coefficient $\beta$ for soft labels is set to 0.3. The models are trained with 2 Quadro RTX 6000 GPUs using a batch size of 64 per GPU. We perform three runs for each setup and obtain the average results.

For student training with the DU Improvement-Set, we randomly replace the batches with samples replayed from the teacher training set to combat catastrophic forgetting \cite{cssl}. For student training with ADS Improvement-Set, the replay dataset is selected as the subset of audio-text pairs that are relevant to the downstream task, namely, the DS Improvement-Set. Note that with the addition of replay, we observed a performance gain of 0.6\%, 0.9\%, and 2.1\% under the ADS curation strategy, on ESC-50, Urbansound8K, and TUT17, respectively compared with the models trained without using replay. We noticed similar trends with other curation strategies as well, so we used replay in all of our results.
\vspace{-.3cm}
\subsection{Downstream Evaluation}
\label{ssec:exp_eval}
We consider the following datasets for downstream evaluation with sound event classification and acoustic scene classification tasks: 
\begin{itemize}
    \setlength\itemsep{.003cm}
    \item ESC-50 \cite{esc} with 2000, 5-second audio clips from 50 environmental sound classes.
    \item UrbanSound8K \cite{us8k} with 8732, 4-second recordings from 10 possible urban sound classes.
    \item TUT Acoustic Scenes 2017 (TUT17) \cite{tut17} with 6300, 10-second clips from 15 possible scene classes.
\end{itemize}
We measure the performance of the models using the accuracy from zero-shot evaluation described in \ref{ssec:method_clap}. Following \cite{clap}, we add a prefix prompt ``\textit{this is a sound of [class label]}'' to each label in the downstream dataset before computing text embeddings.

\section{Results and Discussions}
\label{sec:res}

\subsection{Student Training from Full-dataset Teacher}
\label{ssec:res_fullset}

\begin{table}[]
\caption{Accuracy (percentage) on zero-shot evaluation for teacher and student models based on the teacher trained using full dataset. DU, DS, ADS, and SL denote Domain-Unspecific, Domain-Specific, Augmented Domain-Specific, and soft-labeled loss. Best performances are highlighted in bold.}
\label{tab:fullset}
\centering
\resizebox{7.1cm}{!}{
\begin{tabular}{l|ccc}
 & \multicolumn{3}{c}{\textbf{Zero-Shot Evaluation Set}}\\
\midrule
\textbf{Model} & \textbf{ESC-50} & \textbf{UrbanSound8K} & \textbf{TUT17}\\
\midrule
 CLAP teacher  & 81.9 $\pm$ 0.9 & 74.8 $\pm$ 1.2 & 29.8 $\pm$ 1.3 \\
 \hline 
 SL & 83.1 $\pm$ 1.2 & 73.9 $\pm$ 2.6 & 30.1 $\pm$ 2.1 \\
 \hline
 DU & 82.4 $\pm$ 1.4 & 73.9 $\pm$ 0.2 & 31.5 $\pm$ 1.0 \\
 DU+SL & 83.0 $\pm$ 0.5 &74.9 $\pm$ 1.4 & 29.9 $\pm$ 1.9 \\
 \hline
  DS & 78.8 $\pm$ 0.5 & 73.2 $\pm$ 1.1 & 29.8 $\pm$ 2.6 \\
    DS+SL  & 83.5 $\pm$ 0.6 & 75.5 $\pm$ 1.4 & 31.8 $\pm$ 2.6 \\
    ADS & 84.2 $\pm$ 0.5 & 74.2 $\pm$ 2.1 & 32.5 $\pm$ 1.0 \\
     ADS+SL & \textbf{85.1} $\pm$ 0.7 & \textbf{77.4} $\pm$ 0.6 & \textbf{36.0} $\pm$ 1.8 \\
\end{tabular}
}
\vspace{-.4cm}
\end{table}

In Table \ref{tab:fullset}, we showcase the improvements obtained when the teacher model is trained with the full training set. \footnote{The CLAP model released by Microsoft obtains 82.6\%, 73.4\%, and 29.6\% zero-shot accuracy values as reported in \cite{clap}. Notice that our own CLAP teacher model can closely replicate this performance.}
{When further training the teacher CLAP model with the soft-labeled loss (SL), the accuracy increases by 1.2\% on ESC and by 0.3\% on TUT17 while decreasing by 0.9\% on UrbanSound8K compared to the teacher model, indicating that soft-labeled loss alone cannot yield significant and consistent performance gains.}


We next observe that the domain-unspecific (DU) improvement of the CLAP model improves the zero-shot performance in certain cases. On ESC-50, especially with the addition of soft labels (SL), we are able to increase the zero-shot accuracy from $81.9\%$ to $83.0\%$. However, we do not observe an improvement for all three downstream datasets with DU.


When curating the Improvement-Set with the downstream domain knowledge, we observe that the Domain-Specific (DS) curation alone does not improve the performance: The results validate our hypothesis that learning with hard labels with similar input data from the same batch would adversely impact the quality of learned representations. However, we observe that adding the soft-labeled loss significantly boosts the zero-shot accuracy by exceeding the performance of the teacher model on all three datasets. We finally make the observation that the augmented domain-specific curation along with soft labels (ADS+SL) substantially improves the performance across the board. {The results suggest that additional data with domain-specific curation and soft-labeled loss become more effective when used in conjunction, each of which is crucial to performance improvement.}

\vspace{-.3cm}
\subsection{Student Training from Subset Teacher}
\vspace{-.0cm}
\label{ssec:res_subset}
\begin{table}[]
\caption{Zero-shot accuracy for experiments where the teacher model is trained with 10\% of the original training data. As a reference, we also include the CLAP teacher model trained with the full dataset from Table \ref{tab:fullset}.}
\centering
\resizebox{7.0cm}{!}{
\begin{tabular}{l|ccc}
 & \multicolumn{3}{c}{\textbf{Zero-Shot Evaluation Set}}\\
\midrule
\textbf{Model} & \textbf{ESC-50} & \textbf{UrbanSound8K} & \textbf{TUT17}\\
\midrule
 CLAP teacher (full-dataset) & 81.9 $\pm$ 0.9 & 74.8 $\pm$ 1.2 & 29.8 $\pm$ 1.3 \\
 CLAP teacher (subset) & 74.2 $\pm$ 1.3 & 73.5 $\pm$ 2.0 & 30.9 $\pm$ 1.6 \\
 \hline 
 
DU + SL & 78.9 $\pm$ 0.3 & 73.7 $\pm$ 1.3 & 28.8 $\pm$ 1.1 \\
ADS + SL & \textbf{81.3} $\pm$ 1.0 & \textbf{74.5} $\pm$ 0.7 & \textbf{31.3} $\pm$ 0.5 \\
\end{tabular}
}
\label{tab:res_subset}

\vspace{-.6cm}
\end{table}

In prior experiments, we assume the availability of adequate paired audio and text data for training the teacher model. We would also like to investigate how data curation, label softening, and subsequent student training can be impacted if the amount of paired audio and text is limited during teacher training. To this end, we perform the teacher training by randomly sampling $10\%$ of the original $128\times10^3$ paired audio and caption data. We follow the identical setup for the unsupervised curation and label softening as in the experiments involving the complete dataset. We outline the zero-shot accuracy of the teacher and student models in Table \ref{tab:res_subset}.


Student models trained with the DU Improvement-Set enhance the performance on ESC-50 (from $74.2\%$ to $78.9\%$) and UrbanSound8K (from $73.5\%$ to $73.7\%$) while slightly degrading on TUT17 (from $30.9\%$ to $28.8\%$); the results can be attributed to the domain-unspecific curation, which yields data that are more relevant to environmental sound classes and distinct from acoustic scene recordings. By incorporating domain-specific knowledge, the ADS strategy further enhances the performance of ESC-50 (to $81.3\%$) and UrbanSound8K (to $74.5\%$) beyond that of the teacher model and also exhibits a slight improvement on TUT17 (to $31.3\%$). Quite notably \cem{we observe that with the proposed ADS+SL strategy}, the performance on ESC-50 and UrbanSound8K is similar to that of the teacher model trained with the full dataset. These improvements obtained in the student training demonstrate the effectiveness of the data curation and label softening techniques proposed, even in the absence of ample paired multi-modal data for training the teacher model.

\vspace{-.3cm}

\section{Conclusions}
In this paper, we propose domain-unspecific and domain-specific data curation methods that can effectively improve the zero-shot classification performance of cross-modal representations. We have identified that using domain-specific dataset curation combined with a soft-labeled loss significantly improves the performance over a baseline teacher model. This observation holds even in the case where the baseline teacher is trained with $10\%$ of the original training data. We plan to further improve our approach by incorporating general text corpora into the curation pipeline and exploring more advanced algorithms for audio-text matching in future work.

%



\bibliographystyle{IEEEtran}
\bibliography{refs23}

\begin{thebibliography}{10}
\providecommand{\url}[1]{#1}
\def\UrlFont{\rmfamily}
\providecommand{\newblock}{\relax}
\providecommand{\bibinfo}[2]{#2}
\providecommand\BIBentrySTDinterwordspacing{\spaceskip=0pt\relax}
\providecommand\BIBentryALTinterwordstretchfactor{4}
\providecommand\BIBentryALTinterwordspacing{\spaceskip=\fontdimen2\font plus
\BIBentryALTinterwordstretchfactor\fontdimen3\font minus
  \fontdimen4\font\relax}
\providecommand\BIBforeignlanguage[2]{{%
\expandafter\ifx\csname l@#1\endcsname\relax
\typeout{** WARNING: IEEEtran.bst: No hyphenation pattern has been}%
\typeout{** loaded for the language `#1'. Using the pattern for}%
\typeout{** the default language instead.}%
\else
\language=\csname l@#1\endcsname
\fi
#2}}

\bibitem{gui2023survey}
J.~Gui, T.~Chen, Q.~Cao, Z.~Sun, H.~Luo, and D.~Tao, ``A survey of
  self-supervised learning from multiple perspectives: Algorithms, theory,
  applications and future trends,'' 2023.

\bibitem{simclr}
T.~Chen, S.~Kornblith, M.~Norouzi, and G.~Hinton, ``A simple framework for
  contrastive learning of visual representations,'' in \emph{Proceedings of the
  International Conference on Machine Learning (ICML)}, 2020.

\bibitem{bert}
J.~Devlin, M.-W. Chang, K.~Lee, and K.~Toutanova, ``{BERT}: Pre-training of
  deep bidirectional transformers for language understanding,'' in
  \emph{Proceedings of the Conference of the North {A}merican Chapter of the
  Association for Computational Linguistics}, 2019.

\bibitem{ssast}
Y.~Gong, C.-I. Lai, Y.-A. Chung, and J.~Glass, ``Ssast: Self-supervised audio
  spectrogram transformer,'' in \emph{Proceedings of the AAAI Conference on
  Artificial Intelligence}, vol.~36, no.~10, 2022, pp. 10\,699--10\,709.

\bibitem{clip}
A.~Radford, J.~W. Kim, C.~Hallacy, A.~Ramesh, G.~Goh, S.~Agarwal, G.~Sastry,
  A.~Askell, P.~Mishkin, J.~Clark, G.~Krueger, and I.~Sutskever, ``Learning
  transferable visual models from natural language supervision,'' in
  \emph{Proceedings of the 38th International Conference on Machine Learning},
  2021, pp. 8748--8763.

\bibitem{yuan2021florence}
L.~Yuan, D.~Chen, Y.-L. Chen, N.~C.~F. Codella, X.~Dai, J.~Gao, H.~Hu,
  X.~Huang, B.~Li, C.~Li, C.~Liu, M.~Liu, Z.~Liu, Y.~Lu, Y.~Shi, L.~Wang,
  J.~Wang, B.~Xiao, Z.~Xiao, J.~Yang, M.~Zeng, L.~Zhou, and P.~Zhang,
  ``Florence: A new foundation model for computer vision,'' \emph{ArXiv}, vol.
  abs/2111.11432, 2021.

\bibitem{jia2021scaling}
C.~Jia, Y.~Yang, Y.~Xia, Y.-T. Chen, Z.~Parekh, H.~Pham, Q.~V. Le, Y.~Sung,
  Z.~Li, and T.~Duerig, ``Scaling up visual and vision-language representation
  learning with noisy text supervision,'' 2021.

\bibitem{guzhov2021audioclip}
A.~Guzhov, F.~Raue, J.~Hees, and A.~Dengel, ``Audioclip: Extending clip to
  image, text and audio,'' 2021.

\bibitem{clap}
B.~Elizalde, S.~Deshmukh, M.~A. Ismail, and H.~Wang, ``Clap: Learning audio
  concepts from natural language supervision,'' \emph{arXiv preprint
  arXiv:2206.04769}, 2022.

\bibitem{wu2022wav2clip}
H.-H. Wu, P.~Seetharaman, K.~Kumar, and J.~P. Bello, ``Wav2clip: Learning
  robust audio representations from clip,'' 2022.

\bibitem{vipant}
Y.~Zhao, J.~Hessel, Y.~Yu, X.~Lu, R.~Zellers, and Y.~Choi, ``Connecting the
  dots between audio and text without parallel data through visual knowledge
  transfer,'' in \emph{Proceedings of the 2022 Conference of the North American
  Chapter of the Association for Computational Linguistics: Human Language
  Technologies}.\hskip 1em plus 0.5em minus 0.4em\relax Seattle, United States:
  Association for Computational Linguistics, July 2022, pp. 4492--4507.

\bibitem{laionclap2023}
Y.~Wu*, K.~Chen*, T.~Zhang*, Y.~Hui*, T.~Berg-Kirkpatrick, and S.~Dubnov,
  ``Large-scale contrastive language-audio pretraining with feature fusion and
  keyword-to-caption augmentation,'' in \emph{IEEE International Conference on
  Acoustics, Speech and Signal Processing, ICASSP}, 2023.

\bibitem{audioset}
J.~F. Gemmeke, D.~P.~W. Ellis, D.~Freedman, A.~Jansen, W.~Lawrence, R.~C.
  Moore, M.~Plakal, and M.~Ritter, ``Audio set: An ontology and human-labeled
  dataset for audio events,'' in \emph{Proc. IEEE ICASSP 2017}, New Orleans,
  LA, 2017.

\bibitem{clip_self_distill}
A.~Andonian, S.~Chen, and R.~Hamid, ``Robust cross-modal representation
  learning with progressive self-distillation,'' in \emph{CVPR 2022}, 2022.

\bibitem{softclip}
Y.~Gao, J.~Liu, Z.-H. Xu, T.~Wu, W.~Liu, J.~jin Yang, K.~Li, and X.~Sun,
  ``Softclip: Softer cross-modal alignment makes clip stronger,'' \emph{ArXiv},
  vol. abs/2303.17561, 2023.

\bibitem{FSD50K}
E.~Fonseca, X.~Favory, J.~Pons, F.~Font, and X.~Serra, ``{FSD50K}: an open
  dataset of human-labeled sound events,'' \emph{IEEE/ACM Transactions on
  Audio, Speech, and Language Processing}, vol.~30, pp. 829--852, 2022.

\bibitem{clotho}
K.~Drossos, S.~Lipping, and T.~Virtanen, ``Clotho: an audio captioning
  dataset,'' \emph{ICASSP 2020 - 2020 IEEE International Conference on
  Acoustics, Speech and Signal Processing (ICASSP)}, pp. 736--740, 2019.

\bibitem{audiocaps}
\BIBentryALTinterwordspacing
C.~D. Kim, B.~Kim, H.~Lee, and G.~Kim, ``{A}udio{C}aps: Generating captions for
  audios in the wild,'' in \emph{Proceedings of the 2019 Conference of the
  North {A}merican Chapter of the Association for Computational Linguistics:
  Human Language Technologies, Volume 1 (Long and Short Papers)}.\hskip 1em
  plus 0.5em minus 0.4em\relax Minneapolis, Minnesota: Association for
  Computational Linguistics, June 2019, pp. 119--132. [Online]. Available:
  \url{https://aclanthology.org/N19-1011}
\BIBentrySTDinterwordspacing

\bibitem{macs}
I.~Mart{\'i}n-Morat{\'o} and A.~Mesaros, ``What is the ground truth?
  reliability of multi-annotator data for audio tagging,'' \emph{2021 29th
  European Signal Processing Conference (EUSIPCO)}, pp. 76--80, 2021.

\bibitem{pann}
Q.~Kong, Y.~Cao, T.~Iqbal, Y.~Wang, W.~Wang, and M.~D. Plumbley, ``Panns:
  Large-scale pretrained audio neural networks for audio pattern recognition,''
  \emph{IEEE/ACM Transactions on Audio, Speech, and Language Processing}, 2020.

\bibitem{cssl}
Z.~Wang, C.~Subakan, X.~Jiang, J.~Wu, E.~Tzinis, M.~Ravanelli, and
  P.~Smaragdis, ``Learning representations for new sound classes with continual
  self-supervised learning,'' \emph{IEEE Signal Processing Letters}, vol.~29,
  pp. 2607--2611, 2022.

\bibitem{esc}
\BIBentryALTinterwordspacing
K.~J. Piczak, ``{ESC}: {Dataset} for {Environmental Sound Classification},'' in
  \emph{Proceedings of the 23rd {Annual ACM Conference} on {Multimedia}}.\hskip
  1em plus 0.5em minus 0.4em\relax {ACM Press}, 2015, pp. 1015--1018. [Online].
  Available: \url{http://dl.acm.org/citation.cfm?doid=2733373.2806390}
\BIBentrySTDinterwordspacing

\bibitem{us8k}
J.~Salamon, C.~Jacoby, and J.~P. Bello, ``A dataset and taxonomy for urban
  sound research,'' in \emph{{ACM} International Conference on Multimedia},
  2014.

\bibitem{tut17}
A.~Mesaros, T.~Heittola, A.~Diment, B.~Elizalde, A.~Shah, E.~Vincent, B.~Raj,
  and T.~Virtanen, ``{DCASE} 2017 challenge setup: Tasks, datasets and baseline
  system,'' in \emph{Proceedings of the Detection and Classification of
  Acoustic Scenes and Events 2017 Workshop (DCASE2017)}, November 2017, pp.
  85--92.

\end{thebibliography}
%
%
%
%
%
%
%
%
%

\end{sloppy}
\end{document}